\documentclass[conference]{IEEEtran}
\IEEEoverridecommandlockouts
\usepackage[nolist]{acronym}
\usepackage{cite}
\usepackage{amsmath,amssymb,amsfonts}
\usepackage{algorithmic}
\usepackage{graphicx}
\usepackage{textcomp}
\usepackage{xcolor}
\def\BibTeX{{\rm B\kern-.05em{\sc i\kern-.025em b}\kern-.08em
    T\kern-.1667em\lower.7ex\hbox{E}\kern-.125emX}}
\begin{document}

\title{Comparison of Meta-Heuristics for the Planning of Meshed Power Systems\\
\thanks{The research is part of the project "SpinAI" and funded by the German federal ministry for economic affairs and energy (funding number 0350030B). The authors are solely responsible for the content of this publication.}
}

\author{\IEEEauthorblockN{Florian Sch{\"a}fer}
\IEEEauthorblockA{\textit{Department $e^2n$} \\
\textit{University of Kassel}\\
Kassel, Germany\\
florian.schaefer@uni-kassel.de}
\and
\IEEEauthorblockN{Jan-Hendrik Menke}
\IEEEauthorblockA{\textit{Department $e^2n$} \\
\textit{University of Kassel}\\
Kassel, Germany\\
jan-hendrik.menke@uni-kassel.de}
\and
\IEEEauthorblockN{Martin Braun}
\IEEEauthorblockA{\textit{Department $e^2n$} \\
\textit{University of Kassel, Fraunhofer IEE}\\
Kassel, Germany\\
martin.braun@uni-kassel.de}
}

\maketitle

\begin{abstract}
The power system planning task is a combinatorial optimization problem. The objective function minimizes the economic costs subject to a set of technical and operational constraints. Meta-heuristics are often used as optimization strategies to find solutions to this problem by combining switching, line reinforcement or new line measures. Common heuristics are \ac{GA}, \ac{PSO}, \ac{HC}, \ac{ILS} or newer methods such as \ac{GWO} or \ac{FWA}. In this paper, we compare these algorithms within the same framework. We test each algorithm on 8 different test grids ranging from 73 to 9421 buses. For each grid and algorithm, we start 50 runs with a maximum run time of 1 hour. The results show that the performance of an algorithm depends on the initial grid state, grid size and amount of measures. The \ac{ILS} method is very robust in most cases. In the larger test grids, more exploratory heuristics, e.g., \ac{GA} and \ac{PSO}, find solutions in shorter run times.
\end{abstract}


\section{Introduction}\label{introduction}


The primary objective of power system planning is to meet future demand and generation growth with the restriction of being as reliable, economical and environmental friendly as possible. This is a complex task for several reasons: first, the power system must be designed to cope with all different kinds of loading situations ranging from high load to high generation scenarios. All these loading situations must be taken into account when designing the grid. With the uncertainty of future demand and the growth of variable renewable energy generation, it is not trivial to find long-term investment decisions from an economic point of view. Second, there are many possible alternative measures to be evaluated. These include the reinforcement or replacement of existing lines, switching state optimization or the installation of new line connections between substations. Third, the regulatory frameworks dictate to reduce costs resulting from the operation of the power system. Grid planners need to be able to automatically assess many possible alternatives to find the optimal investment solution.
 
In general, the power system planning task is a combinatorial optimization problem with a single objective function that minimizes the economic costs, resulting from the applied reinforcement/replacement measures, subject to a set of technical and operational constraints \cite{Georgilakis2015}. Technical constraints include adherence to power flow as well as voltage limits. Topological constraints include the supply of all substations and the connection of (distributed) generators.

Meta-heuristic optimization methods are common to solve the planning problem. In this paper, we compare different meta-heuristics and compare these based on eight power system study cases. The different measures considered here to solve the planning problem include upgrading existing lines, switching state optimization and installing additional lines between two substations. 

Most publications test only one specific meta-heuristic for a given task. To be able to compare multiple heuristics a  common framework is needed. It is our goal to compare multiple algorithms in the same framework on multiple power systems. Our work is mainly based on open-source software \cite{Thurner2018,swarmpackagepy, DEAP_JMLR2012,NiaPyJOSS2018} and the framework described in \cite{Scheidler2018}. 

The paper is structured as follows. Sec.\ref{sota} outlines an overview of the state of the art in research. Sec.\ref{basics} explains the planning framework as well as the characteristics of the analyzed heuristics. The 8 power system study cases are described in Sec.\ref{grids}, which are the test cases for the comparison of Sec.\ref{results}. In Sec.\ref{summary} we summarize our findings and give recommendations which algorithms to use.

\section{State of the Art}\label{sota}
Finding the optimal grid structure is the target of a power system planner, who has to assess many alternatives including switching states, reinforced lines, additional connections between substations. The general planning problem is nonlinear and combinatorial with a large number of binary, discrete and continuous variables \cite{Georgilakis2015}. Due to the complexity of the problem, most publications focus on one aspect of the planning problem and optimize single algorithms for the given task. General overviews of different methods are given in \cite{Georgilakis2015,Ganguly2013,Jordehi2015,Resener2018}. 
A similar comparison to this one is \cite{Khan2017}, which compares different heuristics in the same framework.

In general, two approaches are common to solve the planning problem: mathematical programming and heuristic methods. Mathematical programming based methods require to formulate
 a mathematical optimization problem. These problems can then be solved with any MINLP-solver. Since the problem is mixed-integer and non-linear, the solver may not converge for realistic grid sizes. To find (non-optimal) solutions to the problem different relaxation schemes are being developed and, for example, implemented in \cite{Coffrin2018}. Heuristic methods are easier to implement since they do not require the formulation of a mathematical programming model. Instead, they systematically evaluate different measure combinations to solve the problem. Heuristic optimization methods are typically robust and can find near-optimal solutions for complex, large-scale planning problems \cite{Georgilakis2015}. However, these methods cannot guarantee to find the global optimum solution for realistic grid sizes. Well-known and often applied methods are \ac{GA} \cite{Martins2011,Ouyang2010,Zidan2013}, \ac{PSO} \cite{Ziari2012,Olamaei2008}, \ac{HC} \cite{Davidescu2017} or \ac{ILS} \cite{Scheidler2018}. Newer methods are \ac{GWO} \cite{Praveen2018,Sultana2016} or \ac{FWA} \cite{Zhang2018,Imran2014}.

\section{Planning Framework}\label{basics}
\subsection{Optimization Framework}
\label{overview}
The optimization framework we use in this paper is described in detail in  \cite{Scheidler2018}. It is designed to solve the combinatorial power system planning problem with heuristic optimization strategies. The combinatorial optimization problem has the objective (\ref{opt_function}) with constraints (\ref{st_1}) - (\ref{st_5}).

\begin{align}
\underset{s \in 2^M}{\text{minimize}} & & c_m (s) \label{opt_function}\\
\text{subject to} & & st_\text{disconnected}(s) = 0,\label{st_1}\\ 
& & lc_\text{convergence}(s) = 0,\label{st_2}\\ 
& & ll_{(n)}(s) = 0,\label{st_4}\\ 
& & vv_{(n)}(s) = 0,\label{st_5} 
\end{align}

\begin{figure}[!t]
\centering{\includegraphics[width=0.45\textwidth]{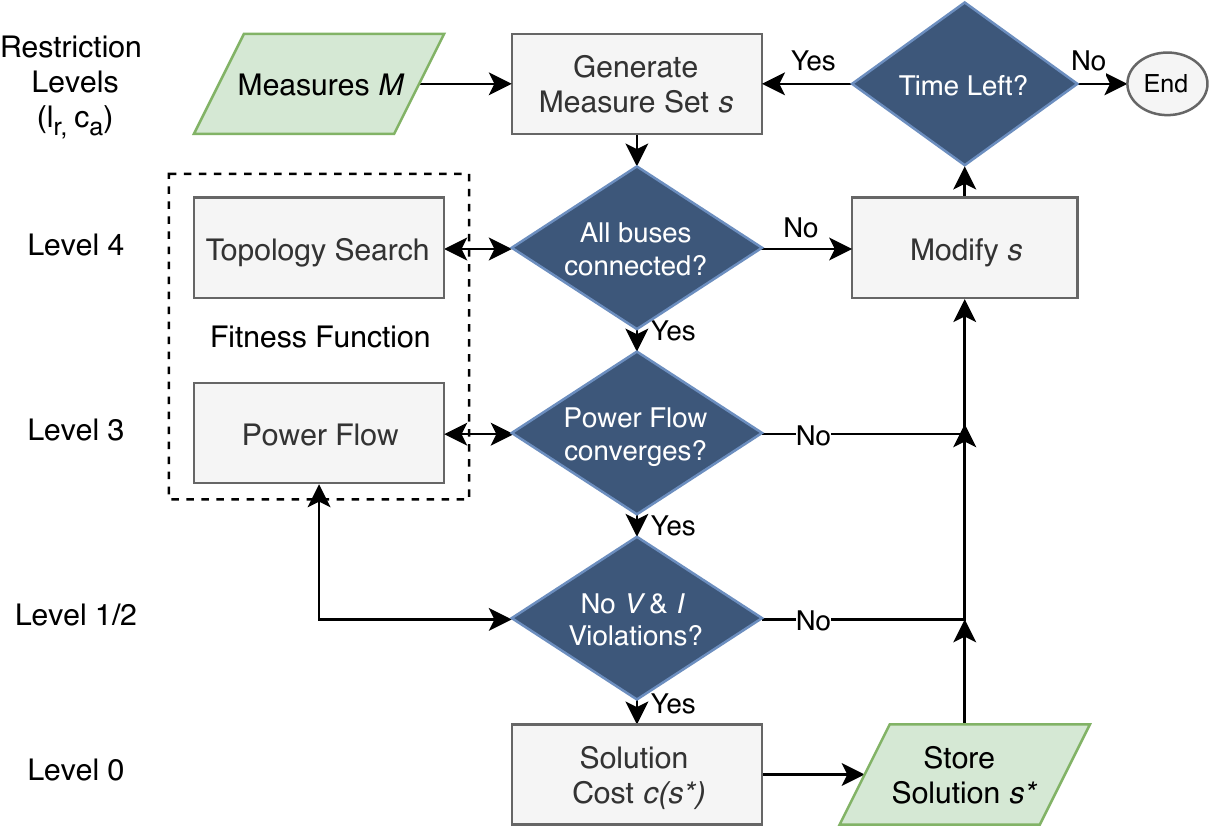}}
\caption{Overview of the planning method from\cite{Scheidler2018} \label{planning_flowchart}}
\end{figure}

The optimization tries to find a measure set $s*$, which does not violate topological or technical constraints and has minimal costs. $s*$ is a subset of all available, pre-defined measures $M$. The framework is based on an iterative solving process. Different \textit{optimization levels} are evaluated, which reflect the severity of the particular violated constraint (\ref{st_1})-(\ref{st_5}) or, if all violations are solved, the height of the investment. These levels are analyzed in a pre-defined order for a given problem. 
Within the optimization, each constraint is formulated by a tuple $(l_r, c_a)$, where the level $l_r$ defines the current violated constraint and $c_a$ equals the optimization cost for the corresponding level. In the first level, it is ensured that all buses are connected to one or multiple reference buses by the connection constraint (\ref{st_1}). If this is the case, a power flow calculation is performed. If it does not converge the convergence constraint (\ref{st_2}) is violated. Otherwise, line loading limits (\ref{st_4}) and voltage limits (\ref{st_5}) are evaluated. If no constraint is violated, the subset $s$ is a solution $s*$ to the planning problem with the cost value $c(s*)$. During the optimization process, this solution is stored and further modified in order to minimize the investment. The optimization terminates either when a time limit or an evaluation limit is reached. Fig.~\ref{planning_flowchart} shows the flow chart of the planning method.

\subsection{Algorithms}\label{algorithms}
\begin{figure}[!ht]
\centering{\includegraphics[width=0.3\textwidth]{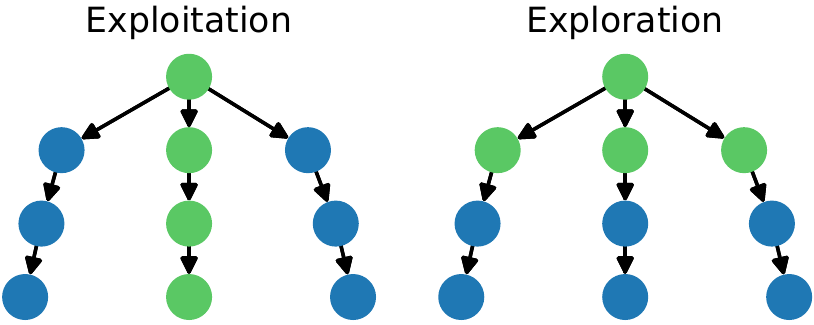}}
\caption{Exploitation and exploration explained on a search tree\label{exploit_explore}}
\end{figure}

The meta-heuristic being compared within the framework include biologically inspired optimizer, such as \ac{GWO}, \ac{GA}, \ac{PSO} as well as other methods, such as \ac{HC} or \ac{ILS} and \ac{FWA}. Every optimization heuristic varies either a single solution or a set of solutions. Depending on the search strategy, we classify the heuristics as ``exploitative'' or ``exploratory''. A high exploration means that the search is ``broad'' and unexplored areas of the solution space are investigated instead of keeping the best-found measures so far. This allows escaping local optima but requires many evaluations. Algorithms with higher exploitation may find ``decent'' solutions with fewer evaluations but may get stuck in local optima more often. Fig.~\ref{exploit_explore} visualizes the optimization process as a search-tree, where every node is a possible set of measures. An algorithm with a high exploitation factor first descends in the graph, shown in Fig.\,\ref{exploit_explore} (left), whereas an algorithm with a high exploration factor first investigates candidates to the left and right of the current candidate (Fig.\,\ref{exploit_explore} right). 
The algorithms we compare are FWA, GWO \cite{swarmpackagepy}, GA \cite{DEAP_JMLR2012}, HC, ILS \cite{Scheidler2018}, PSO \cite{NiaPyJOSS2018}.

\subsection{Measures}
A measure is defined as a single action that can be applied to the grid model to change one property of the grid. To solve the planning problem, each optimization algorithm can choose from a pre-defined set of possible measures for this grid. In the following analysis, three types of measures are considered in the optimization process:

\begin{itemize}
    \item \textbf{REPL} - replacement of existing cables/overhead lines
    \item \textbf{SWITCH} - opening/closing of switches
    \item \textbf{AL} - installation of additional lines between two substations
\end{itemize}

\begin{figure}[!t]
\centering{\includegraphics[width=0.5\textwidth]{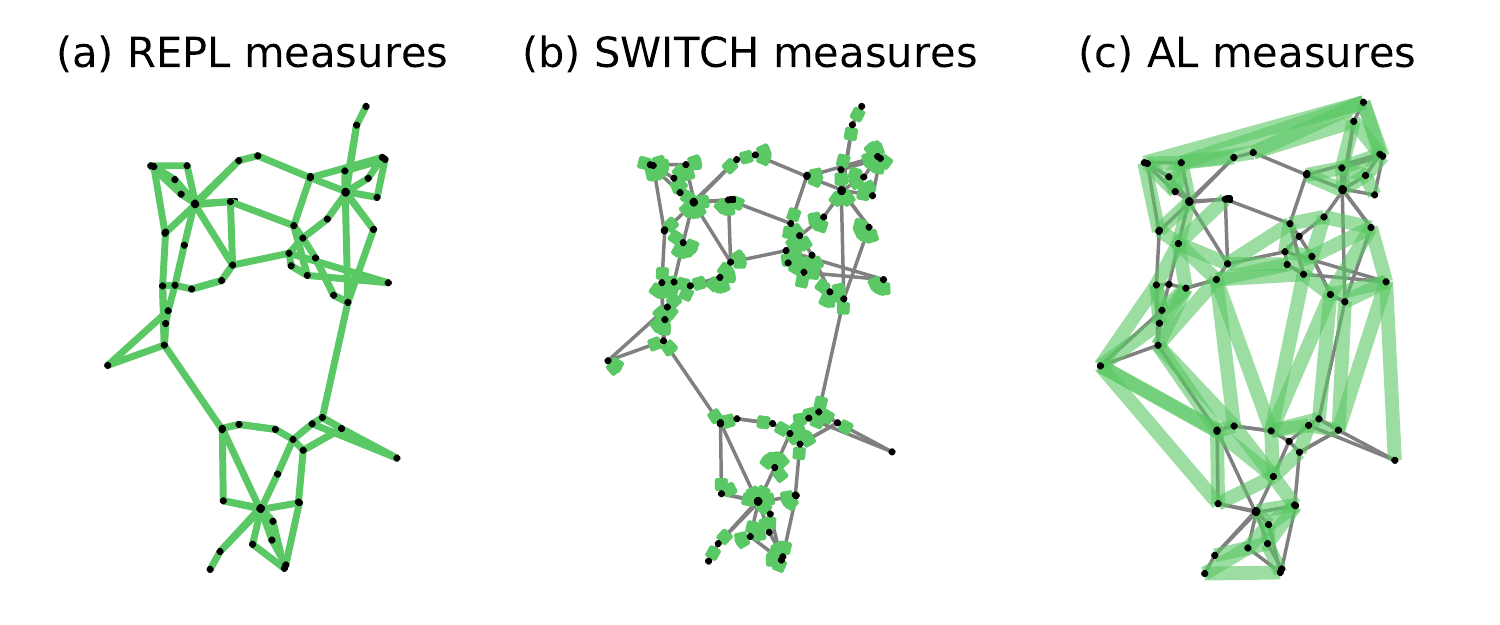}}
\caption{Possible measure types
(a) Replacement (REPL) measures, (b) Switch (SWITCH) measures, (c) Additional Lines (AL) measure\label{measures}}
\end{figure}

For the following analysis, it is assumed that replacing an existing line is equal to installing a parallel line of the same type (doubled admittance). Switches can either be bus-bus switches or bus-line switches. The possible additional line measures are obtained by using a Delaunay triangulation as suggested in \cite{Leon.2018}. Additional line measures are only available as optimization measures if real geographical data for the specific grid is available. Fig.~\ref{measures} shows an example of the possible measure types for the reliability test system grid from \cite{rts}. 

Applying a measure is a binary decision. This results in a combinatorial optimization problem where every set of measures is codified as a binary input vector $m_b \in {m_0, ..., m_N}$ where $m_i = 1$ corresponds to an applied measure and $m_i = 0$ means that the measure is not applied.

\subsection{Load cases}\label{loadcases_chapter}
Load cases are grid states, which define loading situations. Violated grid states have to be solved by the reinforcement measures. Apart from the grid data, provided as a pandapower \cite{Thurner2018} grid, a load case is defined by:
\begin{itemize}
    \item the switching state
    \item outages of assets (lines, transformers)
    \item $P$, $Q$ values of loads
    \item $P$, $Q$ values of (distributed) generators
    \item $V_m$, $\delta$ at slack buses
    \item $P$, $V_m$ at generators
\end{itemize}

Load cases can either be the result of a time series calculation \cite{Schaefer2019} or assumed ``worst case'' situations \cite{Thurner2017}. Load cases result from n-1 loading situations, times of high load or high generation or a combination of these.

\subsection{Modified Cost Function}
\label{cost_function_mod}
The restriction level based design was originally designed for the \ac{HC} algorithm as defined in \cite{Scheidler2018}. \ac{HC} searches in the neighborhood of a solution candidate for an improved candidate. Only one measure is added, removed or replaced at a time. This does not require a comparable and continually improving fitness function across all optimization levels (see Sec.\ref{overview}). The cost of a candidate must be compared only within a restriction level. If a lower level is reached, each candidate at this level has lower optimization costs than a candidate at the upper level. For example, a switching state which results in disconnected buses is worse than a state where all buses are connected, but line loading limits are violated. 

Since the implementations from \cite{swarmpackagepy,DEAP_JMLR2012,NiaPyJOSS2018} need a continuous cost function, which allows comparing solutions based on a single cost value, we normalize the cost function for these algorithms so that each cost evaluation value at one restriction level is between zero and one. The normalized violation levels are defined by: $c_n = l_r + \tanh{c_a}$, with $c_n$, the optimization cost, $c_a$ the level's cost value and the severity $l_r$ of the violated restriction level as explained in \cite{Scheidler2018}, e.g., $l_r = [0, 4]$.

\subsection{Random Spanning Trees}
Each optimization run starts with a population of initial candidates, which are defined as sets of applied measures. These sets include opened switches and optionally pre-defined replacements of lines or new lines. These initial candidates are then varied according to the rules of the optimization method. \ac{HC} and \ac{ILS} start from one candidate and add, remove or replace a single measure in each iteration. All other algorithms start from multiple candidates and can modify multiple measures in one single iteration step. 

GA, PSO, GWO and FWA need initial candidates, which do not violate the topology restrictions. Randomly initialized switching states often result in disconnected areas. To obtain suitable initial conditions, where all buses are connected but switching states differ, we generate multiple different switching states based on random spanning trees generated by the graph-tool library \cite{peixoto_graph-tool_2014}. Similar methods are used in \cite{Najafi2009,Kumar2014}. The spanning trees are subsets of the edges (lines, transformers) of a connected undirected graph that connects all the nodes (buses), without any cycles. For each spanning tree, a power flow calculation is performed. If the power flow calculation converges, the switching state is defined as an initial candidate for the optimization algorithm. If not, $N / 2$ of the remaining opened switches are randomly closed and another power flow calculation is performed. This process is iterated until enough initial candidates have been found.


%

\section{Benchmark Cases}\label{grids}
Eight realistic sized study cases are considered in the comparison. All grid models, except the ``German DSO'' grid, are publicly available from \cite{Thurner2018,rts,simbench}. Characteristics of the benchmark grids are listed in Table \ref{benchmark_cases}. All grid models represent meshed high voltage (110\,kV) or extra-high voltage grids (230\,-\,380\,kV). The number of available line replacement (REPL), switching (SWITCH) and additional line (AL) measures determine the complexity of the planning problem. Each meta-heuristic can choose from $2^M$ combinations of these measures. The SimBench (SB) cases and the RTS case are the only grid models with real available geographical coordinates. These are necessary to determine realistic additional line measures. AL measures were only considered for the RTS case, since we used a pre-released data-set of the SimBench grids in this paper without geodata in the following comparisons. Constraints are defined by upper and lower voltage limits as well as line loading limits (power flow constraints). All buses must be connected within a grid (topology constraint). Optimization constraints are chosen so that all study cases start with violated line loading limits or voltage magnitude violations. For this, line loading and voltage limits are set below the maximum values obtained from a previous power flow calculation. 

\begin{table}[ht]
\renewcommand{\arraystretch}{1.}
\centering
\caption{Overview of analyzed grids and available measures\label{benchmark_cases}}
\label{grid_data}
\begin{tabular}{|l|llllll|}
\hline
Grid    & V & Buses & REPL & SWITCH & AL & Source\\
        & kV & \# &  \#   & \#      & \#  & \\
\hline
SB Mixed & 110 & 328 & 95 & 354 & * & \cite{simbench}\\
SB Urban & 110 & 396 & 113 & 397 & * &  \cite{simbench}\\
German DSO & 110 & 1242 & 110 & 764 & 0 & \\
RTS & 230 & 73 & 37 & 67 & 27 & \cite{rts}\\
case1354pegase & 380 & 1354 & 1751 & 304 & 0 & \cite{Thurner2018}\\
case1888rte & 380 & 1888 & 1976 & 15 & 0 & \cite{Thurner2018}\\
case6515rte & 380 & 6515 & 7421 & 1794 & 0 & \cite{Thurner2018}\\
case9241pegase & 380 & 9241 & 13797 & 2901 & 0 & \cite{Thurner2018}\\
\hline
\end{tabular}
\end{table}

\section{Results}\label{results}
\subsection{Algorithm Comparison}
All heuristic optimizer depend on the initial candidates (measure sets) and random variations of these. To reduce the impact of randomness in the comparison, we run each algorithm 50 times for each grid with a limited time of 1 hour per run. For 6 algorithms and 8 grids, this results in 2400 runs calculated on a high-performance cluster with 48 cores. Each algorithm receives the same input data, which consists of the power system data, the available measures, and a set of load cases to solve.

We want to compare the performance of each algorithm by different criteria:
\begin{itemize}
    \item the time needed until the best solution is found within each run
    \item the share of best solutions found in all runs for each algorithm
    \item the time spent within each kind of evaluation type (topology, power flow, cost)
\end{itemize}

\subsubsection{Example Run}
\begin{figure}[!t]
\centering{\includegraphics[width=0.5\textwidth]{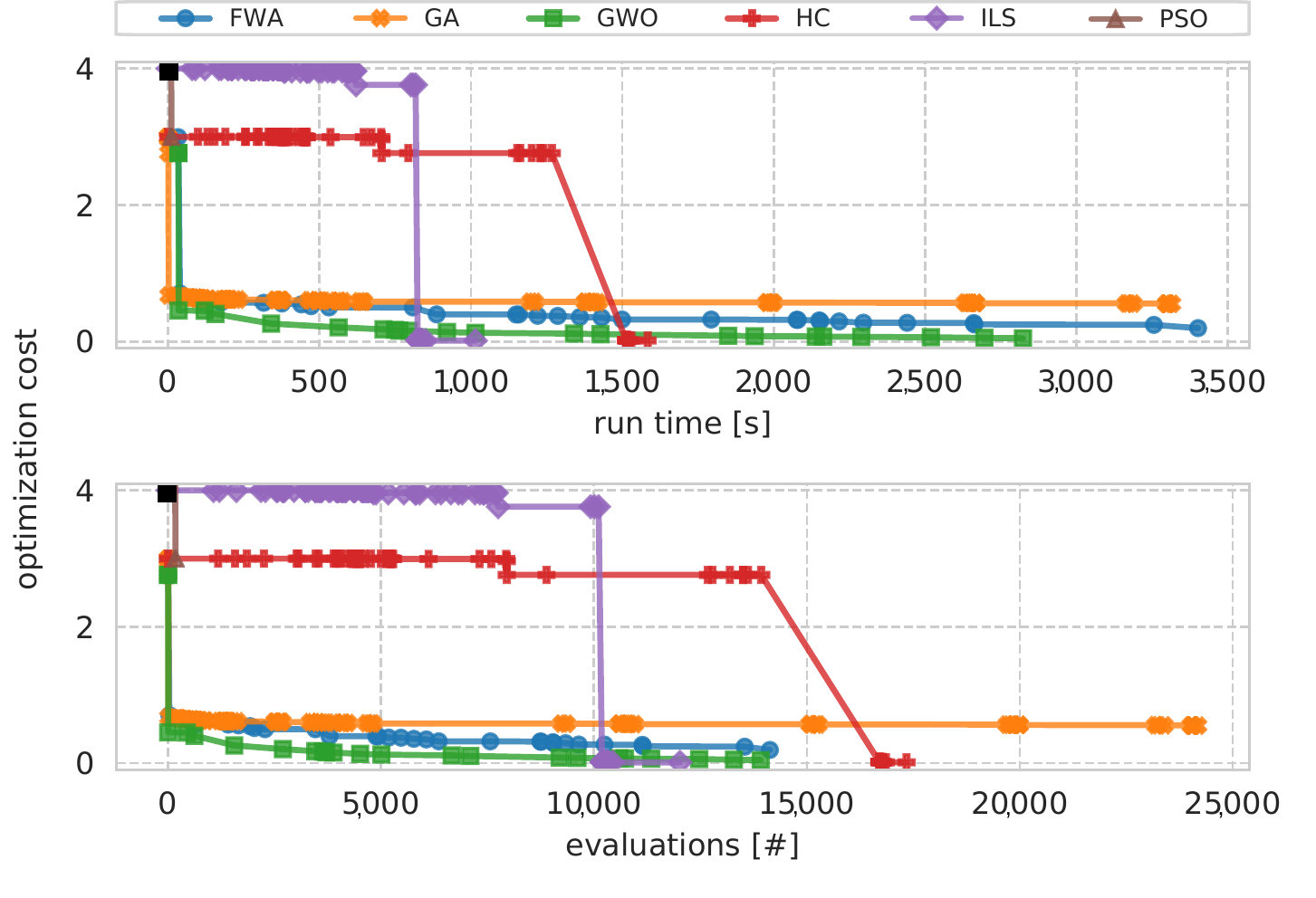}}
\caption{Exemplary improvement of all algorithms for the case1354pegase by time (top) and evaluations (bottom) \label{example_run}}
\end{figure}

Fig.~\ref{example_run} shows the improvement of the best-found solution over time (a) and evaluations (b) for the case1354pegase case. It can be seen that all algorithms found local optima at some point and do not further improve within the 1 hour run time. \ac{HC} and \ac{ILS} found local optima after 1700 or 1200\,s. Before that time, only a slight improvement within the voltage violation level can be seen. All other algorithms, except \ac{PSO}, find a valid solution without violated restrictions in less than 50\,s. \ac{GWO}, \ac{FWA} and \ac{GA} continually improve with longer run times. The figure shows that \ac{GA} performs many more evaluations in the same run time. This is since the heuristic evaluates many switching states which result in disconnected buses. The graph search, necessary for this topology evaluation, is much faster than a power flow calculation. This leads to more unsuccessful evaluations per time.

\subsubsection{Overview of Best Solutions Found}
\begin{figure}[!t]
\centering{\includegraphics[width=0.5\textwidth]{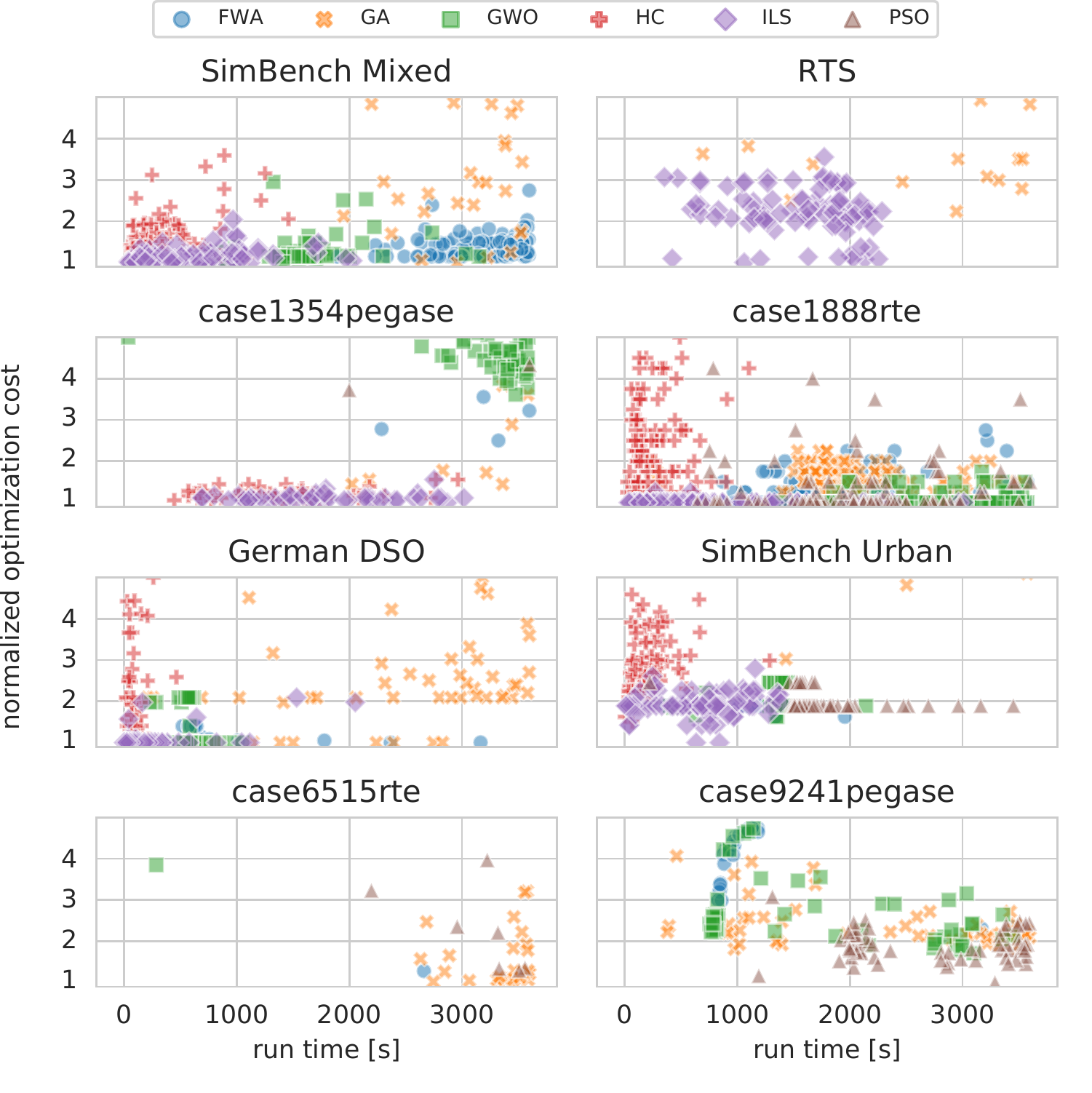}}
\caption{Best result found of 50 runs for each algorithm and grid with a maximum run time of 1 hour.\label{combined_all}}
\end{figure}

Fig.~\ref{combined_all} shows the results for 50 runs per algorithm and grid. The maximum run time is limited to 1 hour. All solutions are normalized by dividing the best-found solution by the total best-found solution in all runs. Results which have costs higher than 5x the costs of the best solutions are not shown. 

\textit{SB Mixed - }
The \ac{HC} algorithm can find solutions with normalized costs between 1 and 2 in most runs. In some runs, the normalized costs are higher than 2 and need longer run times to converge. These solutions are local optima found by the \ac{HC} algorithm. \ac{ILS} escapes these local optima with fewer additional evaluations and finds solutions with costs lower than 2 for this grid. \ac{GWO} and \ac{FWA} can find solutions with costs less than 2 in most cases, but need longer run times. The \ac{PSO} is able to find results with costs lower than 5 in none of the runs. The \ac{GA} solutions have the highest spread of all algorithms. Only some of the solutions have normalized costs less than 2. 

\textit{RTS - }
The result shown in Fig.~\ref{combined_all} that only \ac{ILS} and \ac{GA} were able to find solutions with normalized costs less than 5 in all runs. All other algorithms do not converge within the one-hour time limit in all runs.

\textit{case1354pegase - }
In most of the 50 runs, only \ac{HC}, \ac{ILS} and \ac{GA} are able to find solutions with normalized costs less than 2. The \ac{GWO} and \ac{FWA} are able to find solutions with normalized costs less than 5x the costs of the solutions found by \ac{HC} or \ac{ILS} in run times longer than 3000\,s.

\textit{case1888rte - }
All algorithms are able to find solutions with normalized costs less than 2 in the case1888rte grid. \ac{GA}, \ac{FWA} and \ac{HC} solution costs spread between 1 and 5. The algorithms get stuck in local optima. \ac{ILS} often finds the results with the lowest costs in the shortest run time.

\textit{German DSO - }
In this grid \ac{ILS} shows very good performance. It often finds low-cost solutions within a shorter run time than the other algorithms. \ac{GA} has a high spread in the solution cost. The \ac{PSO} and \ac{FWA} barely find solutions at all.

\textit{SB Urban - }
In the SB Urban case, \ac{ILS} also outperforms all other algorithms in terms of the best solution found in short run times. \ac{GA} is seldom able to find solutions with low costs.

\textit{case6515rte - }
Results of the case6515rte grids show that the initialization with random spanning trees and the high exploration factor is more beneficial for this grid. The \ac{GA} finds most of the solutions with the lowest costs. All other algorithms could not find solutions with similar low costs (except for two outliers). For this grid, \ac{HC} and \ac{ILS} were not able to find a single set of measures, which did not violate the line loading restrictions within the one hour run time limit. 

\textit{case9241pegase - }
In the largest grid, case9241pegase, \ac{ILS}, and \ac{HC} could not find a single solution without violated restrictions. Since only one measure is applied at the time, the search only gradually improves, while the exploratory algorithms are able to improve faster. \ac{GA} and \ac{PSO} show good results for this grid.

\subsubsection{Best Solution Found After the Run Time Limit}
\begin{figure}[!t]
\centering{\includegraphics[width=0.5\textwidth]{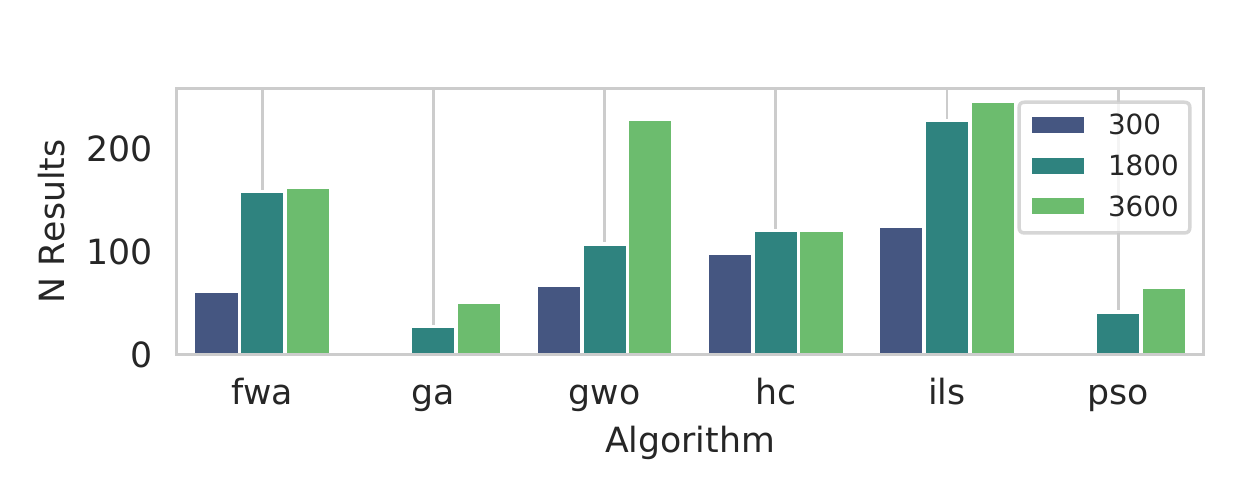}}
\caption{Number of best results found after 300\,s, 1800\,s and 1\,hour. Most algorithms improve with longer run times. ILS and GWO find most often the best results during the run time limit. \label{resbest}}
\end{figure}

Fig.~\ref{resbest} shows a comparison of the total number of best-found solutions for all algorithms after the one hour run time limit. This shows how each algorithm improves with longer run times. All algorithms improve with longer run times between 300 and 1800\,s. \ac{HC} does not improve after 1800\,s and \ac{FWA} improves slightly. \ac{ILS}, \ac{GWO}, \ac{GA} and \ac{PSO} find better solutions after 1800\,s. Especially for the largest grid (case9241pegase) \ac{GA} and \ac{PSO} improve with longer run times. \ac{ILS} and \ac{GWO} have most often found the best solutions of all algorithms within the run time limit. \ac{ILS} additionally improved faster than \ac{GWO}.

\subsubsection{Evaluations by Type}
\begin{figure}[!t]
\centering{\includegraphics[width=0.5\textwidth]{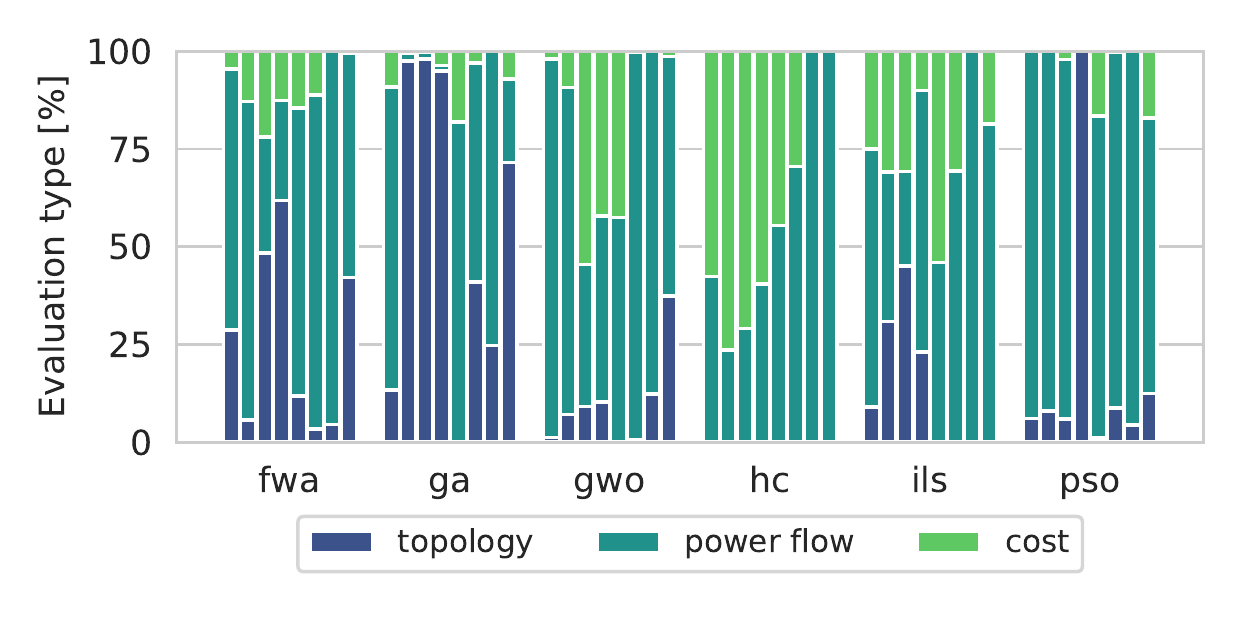}}
\caption{Percentage of time spent within an optimization level. Topology violations result from invalid switching states, power flow evaluations are used to evaluate line loading and voltage limits, cost evaluations determine costs when no constraints are violated.\label{evalsbar}}
\end{figure}

Depending on an algorithm's search strategy (exploitation vs. exploration), either topology or power flow evaluations are performed more often. Topology evaluations need significantly less computational time than power flow evaluations with up to a factor of 100. We analyze the time of each algorithm spent by the evaluation type in Fig.~\ref{evalsbar}. It shows the relative time spent by each algorithm with either evaluating topology restrictions, technical restrictions (line loading and voltage evaluation) and cost evaluation. \ac{HC} and \ac{ILS} perform mostly power flow and cost evaluations in all grids. This is because only one measure is applied at a time, which seldom leads to topology violations (disconnected buses). The exploratory algorithms, such as \ac{GA}, have a much higher share of topology evaluations in comparison to the exploitative strategies. This allows them to find switching states with lower line loadings and voltage deviations but results in many topology evaluations. 

\section{Summary and Recommendations}\label{summary}
In this paper, we compared 6 meta-heuristics for 8 power system study cases. The comparison showed that no algorithm performs ``best'' for all of the analyzed cases. All algorithms got stuck in local optima in some runs due to their stochastic approach of searching. Algorithms with higher exploratory behavior can find ``decent'' solutions faster than exploitative ones. However, as soon as a solution without violations is found, the exploitative algorithms improve in a shorter time. The following conclusions can be summarized:

\begin{itemize}
    \item The exploitative algorithms, \ac{HC}, and \ac{ILS} are most robust and performed best in most grids except for the two largest ones. These algorithms often need longer run times to find the first solution.
    \item The exploratory algorithms, \ac{GA}, and \ac{PSO} showed a good performance when choosing from a large number of measures. Shorter run times were needed to find a first solution. Many grids states are evaluated with topology violations without significant improvement.
    \item \ac{GWO}, and \ac{FWA} showed a trade-off between solution quality and run time.
    \item Initial solutions are very relevant for convergence. The random spanning tree method helped to find valid initial switching states.
\end{itemize}

Based on our findings we conclude that the \ac{ILS} heuristic from \cite{Scheidler2018} is most robust for the given optimization task. We assume several reasons for this: First, \ac{HC} and \ac{ILS} heuristics change a single switching state at a time. In comparison, the exploratory algorithms often resulted in invalid switching states. Second, the \ac{HC} and \ac{ILS} cost function must not be normalized as explained in Section\,\ref{cost_function_mod}. Further research is needed to verify the benefit of different cost functions. Third, the combined optimization of all three measure types favors greedy algorithms. A comparison is needed with only single measure type optimization, e.g., line replacement optimization for a fixed topology. Additionally, different load cases for the benchmark grids should be compared in a sensitivity analysis.

\bibliographystyle{IEEEtran}
\bibliography{conference_041818}

\begin{thebibliography}{10}
\providecommand{\url}[1]{#1}
\csname url@samestyle\endcsname
\providecommand{\newblock}{\relax}
\providecommand{\bibinfo}[2]{#2}
\providecommand{\BIBentrySTDinterwordspacing}{\spaceskip=0pt\relax}
\providecommand{\BIBentryALTinterwordstretchfactor}{4}
\providecommand{\BIBentryALTinterwordspacing}{\spaceskip=\fontdimen2\font plus
\BIBentryALTinterwordstretchfactor\fontdimen3\font minus
  \fontdimen4\font\relax}
\providecommand{\BIBforeignlanguage}[2]{{%
\expandafter\ifx\csname l@#1\endcsname\relax
\typeout{** WARNING: IEEEtran.bst: No hyphenation pattern has been}%
\typeout{** loaded for the language `#1'. Using the pattern for}%
\typeout{** the default language instead.}%
\else
\language=\csname l@#1\endcsname
\fi
#2}}
\providecommand{\BIBdecl}{\relax}
\BIBdecl

\bibitem{Georgilakis2015}
P.~S. Georgilakis and N.~D. Hatziargyriou, ``A review of power distribution
  planning in the modern power systems era: Models, methods and future
  research,'' \emph{Electric Power Systems Research}, vol. 121, pp. 89--100,
  apr 2015.

\bibitem{Thurner2018}
L.~Thurner, A.~Scheidler, F.~Schafer, J.-H. Menke, J.~Dollichon, F.~Meier,
  S.~Meinecke, and M.~Braun, ``Pandapower{\textemdash}an open-source python
  tool for convenient modeling, analysis, and optimization of electric power
  systems,'' \emph{{IEEE} Transactions on Power Systems}, vol.~33, no.~6, pp.
  6510--6521, nov 2018.

\bibitem{swarmpackagepy}
``Swarmpackagepy,'' {github.com/SISDevelop/SwarmPackagePy}, accessed:
  2019-10-16.

\bibitem{DEAP_JMLR2012}
F.-A. Fortin, F.-M. {De Rainville}, M.-A. Gardner, M.~Parizeau, and C.~Gagn\'e,
  ``{DEAP}: Evolutionary algorithms made easy,'' \emph{Journal of Machine
  Learning Research}, vol.~13, pp. 2171--2175, jul 2012.

\bibitem{NiaPyJOSS2018}
G.~Vrban{\v{c}}i{\v{c}}, L.~Brezo{\v{c}}nik, U.~Mlakar, D.~Fister, and
  I.~{Fister Jr.}, ``{NiaPy: Python microframework for building nature-inspired
  algorithms},'' \emph{{Journal of Open Source Software}}, vol.~3, 2018.

\bibitem{Scheidler2018}
A.~Scheidler, L.~Thurner, and M.~Braun, ``Heuristic optimisation for automated
  distribution system planning in network integration studies,'' \emph{{IET}
  Renewable Power Generation}, vol.~12, no.~5, pp. 530--538, apr 2018.

\bibitem{Ganguly2013}
S.~Ganguly, N.~C. Sahoo, and D.~Das, ``Recent advances on power distribution
  system planning: a state-of-the-art survey,'' \emph{Energy Systems}, vol.~4,
  no.~2, pp. 165--193, jan 2013.

\bibitem{Jordehi2015}
A.~R. Jordehi, ``Optimisation of electric distribution systems: A review,''
  \emph{Renewable and Sustainable Energy Reviews}, vol.~51, pp. 1088--1100, nov
  2015.

\bibitem{Resener2018}
M.~Resener, S.~Haffner, L.~A. Pereira, and P.~M. Pardalos, ``Optimization
  techniques applied to planning of electric power distribution systems: a
  bibliographic survey,'' \emph{Energy Systems}, vol.~9, no.~3, pp. 473--509,
  jan 2018.

\bibitem{Khan2017}
B.~Khan and P.~Singh, ``Selecting a meta-heuristic technique for smart
  micro-grid optimization problem: A comprehensive analysis,'' \emph{{IEEE}
  Access}, vol.~5, pp. 13\,951--13\,977, 2017.

\bibitem{Coffrin2018}
C.~Coffrin, R.~Bent, K.~Sundar, Y.~Ng, and M.~Lubin, ``{PowerModels}. {JL}: An
  open-source framework for exploring power flow formulations,'' in \emph{2018
  Power Systems Computation Conference ({PSCC})}.\hskip 1em plus 0.5em minus
  0.4em\relax {IEEE}, jun 2018.

\bibitem{Martins2011}
V.~F. Martins and C.~L.~T. Borges, ``Active distribution network integrated
  planning incorporating distributed generation and load response
  uncertainties,'' \emph{{IEEE} Transactions on Power Systems}, vol.~26, no.~4,
  pp. 2164--2172, nov 2011.

\bibitem{Ouyang2010}
W.~Ouyang, H.~Cheng, X.~Zhang, L.~Yao, and M.~Bazargan, ``Distribution network
  planning considering distributed generation by genetic algorithm combined
  with graph theory,'' \emph{Electric Power Components and Systems}, vol.~38,
  no.~3, pp. 325--339, jan 2010.

\bibitem{Zidan2013}
A.~Zidan, M.~F. Shaaban, and E.~F. El-Saadany, ``Long-term multi-objective
  distribution network planning by {DG} allocation and feeders'
  reconfiguration,'' \emph{Electric Power Systems Research}, vol. 105, pp.
  95--104, dec 2013.

\bibitem{Ziari2012}
I.~Ziari, G.~Ledwich, A.~Ghosh, and G.~Platt, ``Integrated distribution systems
  planning to improve reliability under load growth,'' \emph{{IEEE}
  Transactions on Power Delivery}, vol.~27, no.~2, pp. 757--765, apr 2012.

\bibitem{Olamaei2008}
J.~Olamaei, T.~Niknam, and G.~Gharehpetian, ``Application of particle swarm
  optimization for distribution feeder reconfiguration considering distributed
  generators,'' \emph{Applied Mathematics and Computation}, vol. 201, no. 1-2,
  pp. 575--586, jul 2008.

\bibitem{Davidescu2017}
G.~Davidescu and V.~Vyatkin, ``Network planning and self-repair in models of
  urban distribution networks via hill climbing,'' in \emph{{IECON} 2017 - 43rd
  Annual Conference of the {IEEE} Industrial Electronics Society}.\hskip 1em
  plus 0.5em minus 0.4em\relax {IEEE}, oct 2017.

\bibitem{Praveen2018}
P.~Praveen, S.~Ray, J.~Dasl, and A.~Bhattacharya, ``Multi-objective power
  system expansion planning with renewable intermittency and considering
  reliability,'' in \emph{2018 Internat2018 International Conference on
  Computation of Power, Energy, Information and Communication ({ICCPEIC})ional
  conference on computation of power, energy, Information and Communication
  ({ICCPEIC})}.\hskip 1em plus 0.5em minus 0.4em\relax {IEEE}, mar 2018.

\bibitem{Sultana2016}
U.~Sultana, A.~B. Khairuddin, A.~Mokhtar, N.~Zareen, and B.~Sultana, ``Grey
  wolf optimizer based placement and sizing of multiple distributed generation
  in the distribution system,'' \emph{Energy}, vol. 111, pp. 525--536, sep
  2016.

\bibitem{Zhang2018}
Y.~Zhang, J.~Liu, H.~Zhou, K.~Guo, and F.~Tang, ``Intelligent reconfiguration
  for distributed power network with multivariable renewable generation,'' in
  \emph{2018 Asian Conference on Energy, Power and Transportation
  Electrification ({ACEPT})}.\hskip 1em plus 0.5em minus 0.4em\relax {IEEE},
  oct 2018.

\bibitem{Imran2014}
A.~M. Imran, M.~Kowsalya, and D.~Kothari, ``A novel integration technique for
  optimal network reconfiguration and distributed generation placement in power
  distribution networks,'' \emph{International Journal of Electrical Power {\&}
  Energy Systems}, vol.~63, pp. 461--472, dec 2014.

\bibitem{Leon.2018}
T.~Leon, \emph{Structural Optimizations in Strategic Medium Voltage Power
  System Planning}, ser. Energy Management and Power System Operation.\hskip
  1em plus 0.5em minus 0.4em\relax Kassel, Hess: {Kassel University Press},
  2018, vol.~4.

\bibitem{rts}
``Reliability test system - grid modernization lab consortium,''
  {github.com/GridMod/RTS-GMLC}, accessed: 2019-10-16.

\bibitem{Schaefer2019}
F.~M.~B. F.~Schaefer, J.-H.~Menke, ``Time series based power system planning
  including storage systems and curtailment strategies,'' in \emph{CIRED, 25th
  International Conference on Electricity Distribution, Madrid}.\hskip 1em plus
  0.5em minus 0.4em\relax {CIRED}, jun 2019.

\bibitem{Thurner2017}
L.~Thurner, A.~Scheidler, A.~Probst, and M.~Braun, ``Heuristic optimisation for
  network restoration and expansion in compliance with the single-contingency
  policy,'' \emph{{IET} Generation, Transmission {\&} Distribution}, vol.~11,
  no.~17, pp. 4264--4273, nov 2017.

\bibitem{peixoto_graph-tool_2014}
T.~P. Peixoto, ``The graph-tool python library,'' \emph{figshare}, 2014.

\bibitem{Najafi2009}
S.~Najafi, S.~Hosseinian, M.~Abedi, A.~Vahidnia, and S.~Abachezadeh, ``A
  framework for optimal planning in large distribution networks,'' \emph{{IEEE}
  Transactions on Power Systems}, vol.~24, no.~2, pp. 1019--1028, may 2009.

\bibitem{Kumar2014}
D.~Kumar, S.~Samantaray, and G.~Joos, ``A reliability assessment based graph
  theoretical approach for feeder routing in power distribution networks
  including distributed generations,'' \emph{International Journal of
  Electrical Power {\&} Energy Systems}, vol.~57, pp. 11--30, may 2014.

\bibitem{simbench}
``Simbench - benchmark data set for grid analysis, grid planning and grid
  operation management,'' {https://simbench.de/en}, accessed: 2019-10-16.

\end{thebibliography}

\begin{acronym}
\acro{BMWI}{Bundes-Ministerium f{\"u}r Wirtschaft und Energie}
\acro{CAPEX}{capital expenditures}
\acro{DER}{distributed energy resource}
\acro{DG}{distributed generator}
\acro{DSM}{demand side management}
\acro{DSO}{distribution system operator}
\acro{FWA}{fireworks algorithm}
\acro{GA}{genetic algorithm}
\acro{GWO}{grey wolf optimizer}
\acro{HC}{hill climbing}
\acro{ICT}{information and communication technology}
\acro{ILS}{iterated local search}
\acro{IT}{information technology}
\acro{MILP}{Mixed Integer Linear Programming}
\acro{MINLP}{Mixed Integer Non-Linear Programming}
\acro{OPF}{optimal power flow}
\acro{OPEX}{operating expenditures}
\acro{PDP}{power distribution planning}
\acro{PSO}{particle swarm optimization}
\acro{PSP}{Power System Planning}
\acro{RES}{renewable energy resource}
\acro{SCADA}{supervisory control and data acquisition}
\acro{SCP}{single contingency policy}
\end{acronym}

\end{document}